\documentclass{article}

\usepackage{amsmath}
\usepackage{amscd}
\usepackage{amsthm}
\usepackage{amssymb} \usepackage{latexsym}
\usepackage{eufrak}
\usepackage{euscript}
\usepackage{epsfig}
\usepackage{graphics}
\usepackage{array}
\usepackage{enumerate}

\theoremstyle{theorem}

\theoremstyle{corollary}

\theoremstyle{lemma}

\theoremstyle{definition}

\theoremstyle{proof}

\theoremstyle{remark}

\newcommand{\bel}[1]{\begin{equation}\label{#1}}

\newcommand{\be}{\begin{equation}}

\newcommand{\ba}{\begin{eqnarray}}
\newcommand{\ea}{\end{eqnarray}}
\newcommand{\rf}[1]{(\ref{#1})}
\newcommand{\bi}{\bibitem}
\newcommand{\qe}{\end{equation}}

\newcommand{\N}{\mathbb{N}}

\begin{document}
\title{Spectral plots and the representation and interpretation of
  biological data}
\author{Anirban Banerjee, J\"urgen Jost \\[2ex]
 {\small Max Planck Institute for Mathematics in the
  Sciences,}\\ 
{\small Inselstr.22, 04103 Leipzig, Germany,}\\
 {\tiny banerjee@mis.mpg.de,
  jost@mis.mpg.de}}
\maketitle
\begin{abstract}
It is basic question in biology and other fields to identify the
characteristic properties that on one hand are shared by structures from a
particular realm, like gene regulation, protein-protein interaction or
neural networks or foodwebs, and that on the other hand distinguish them
from other structures.
We introduce and apply a general method, based on the spectrum of the
normalized graph Laplacian, that yields representations, the spectral plots,
that allow us to find and visualize such properties systematically. We
present such visualizations for a wide range of  biological networks and
compare them with those for networks derived from theoretical schemes. The
differences that we find are quite striking and suggest that the search for
universal properties of biological networks should be complemented by an
understanding of more specific features of biological organization
principles at different scales.
\end{abstract}
\bigskip

\section{Introduction}
The large volume of typical data sets produced in modern molecular,
cell and neurobiology raises certain systematic questions, or, more
precisely, brings new aspects to some old scientific issues. These, or
at least the ones we wish to address in this note, are:
\begin{enumerate}
\item Given a particular biological structure, which of each features
  or qualities are universal, that is, shared by other structures
  within a certain class, and what is unique and special for the
  structure at hand?
\item Given a large and complex structure, should we focus on
  particular and specific aspects and quantities in detail, or should
  we try to obtain, at least at some coarse level, a simultaneous
  representation of all its qualitative features?
\end{enumerate}
Clearly, these questions can be posed, but not answered in such
generality. Here, we look at those issues for a particular type  of
 biological data, namely those that are presented as graphs or
 networks. For these, we introduce and discuss a certain
 representation, a spectral plot, that allows for analyzing the
 questions raised above, in both cases with an emphasis on the second
 alternative.

\section{Graphs and their invariants} 
Many biological data sets are, or can be, represented as networks,
that is, in terms of the formal structure of a 
graph. The vertices of a graph stand for the units in question, like
genes, proteins, cells, neurons, and an edge between vertices
expresses some correlation or interaction between the corresponding
units. These edges can be directed, to encode the direction of
interaction, for example via a synaptic connection between neurons,
and weighted, to express the strength of interaction, like a synaptic
weight. Here, for simplicity of presentation, we only consider the
simplest type of a graph, the undirected and unweighted one, although
our methods apply and 
our considerations remain valid in the general situation. Thus, an
edge expresses the presence of some interaction, connection or direct
correlation between two vertices, regardless of its direction or
strength. Clearly, this abstraction may loose many important details,
but we are concerned here with what it preserves. Thus, we shall
investigate the two issues raised in the introduction for, possibly
quite large, graphs.\\\\
An (unweighted, undirected) graph is given by a set $V$ of vertices
$i,j,\dots $ and a set of edges $E$ which are simply unordered pairs
$(i,j)$ of  vertices.\footnote{Some general references for graph theory
  are \cite{Bol1,GoRo}.} (Usually, one assumes that there can only be
edges between different vertices, that is, there are no self-loops
from a vertex to itself.) Thus, when the pair $(i,j)$ is in $E$, then
the vertices $i$ and $j$ are connected by an edge, and we call
them neighbors and write this relation as $i\sim j$. The degree $n_i$
of a vertex $i$ is defined to be the number of its
neighbors. Already for rather modest numbers of vertices, say 20, the
number of different graphs is bewilderingly high.\footnote{As always in mathematics, there is a notion of
  isomorphism: Graphs $\Gamma_1$ and $\Gamma_2$ are called isomorphic when
  there is a one-to-one map $\rho$ between the vertices of $\Gamma_1$
  and $\Gamma_2$ that preserves the neighborhood relationship, that is
  $i\sim j$ precisely if $\rho(i)\sim \rho(j)$. Isomorphic graphs are
  considered to be the same because they cannot be distinguished by
  their properties. In other words, when we speak about different
  graphs, we mean non-isomorphic ones.}
Thus, it becomes impractical, if not impossible, to list all different
graphs with a given number of vertices, unless that number is rather
small. Also, drawing a graph with a large number of vertices is not
helpful for visual analysis because the structure will just look
convoluted and complicated instead of transparent. On the other hand,
graphs can be qualitatively quite different, and understanding this is
obviously crucial for the analysis of the represented biological
structures. For example, the maximal distance (number of edges) between two vertices in
a graph of size $N$ can vary between $1$ and $N-1$, depending on the
particular graph. When the graph is complete, that is, every vertex
is connected with every other one, any two vertices have the distance
1, whereas for a chain where vertex $i_1$ is only connected to $i_2$,
$i_2$ then also to $i_3$, and so on, the first and the last vertex
have distance $N-1$. For most graphs, of course, some intermediate
value will be realized, and one knows from the theory of random
graphs that for a typical graph this maximal distance is of the order
$\log N$. So, this maximal distance is one graph invariant, but still,
rather different graphs can have the same value of this
invariant. Adjoining a long sidechain to a complete subgraph can
produce the same value as an everywhere loosely connected, but rather
homogeneous graph. The question then emerges whether one should look
for other, additional or more comprehensive, invariants, or whether one
should adopt an entirely different strategy for capturing the
essential properties of some given graph.\\
In fact, there are many graph invariants that each capture certain
important qualitative aspects, and that have been extensively studied
in graph theory (see e.g. \cite{Bol1,GoRo}). These range from rather simple and
obvious ones, like maximal or average degree of vertices or distance
between them, to ones that reflect more global aspects, like how
difficult it is to separate the graph into disjoint components (see
e.g. \cite{Chung}),
commmunities (e.g.\cite{New}) or classes, or to
synchronize coupled dynamics operating at the individual
vertices (e.g. \cite{JJ1}). For the sake of the subsequent discussion, we call these
properties cohesion and coherence, resp. Recently, the degree distribution of the vertices received
much attention. Here, for each $n \in \N$, one lists the number $k_n$
vertices $i$
 in the graph with degree $n_i=n$ and then looks at the behavior of
 $k_n$ as a function of $n$. In random graphs as introduced by Erd\"os
 and R\'enyi (e.g. \cite{Bol2}), that is, graphs where
 one starts with a given collection of, say $N$, vertices and to each
 pair $i,j$ of vertices, one assigns an edge with some probability
 $0<p<1$, typically $k_n \sim e^{-\sigma n}$ for some constant
 $\sigma$, that is the degree sequence decays
 exponentially.\footnote{In more precise terms, the degrees are
   Poisson distributed in the limit of an infinite graph size.} Barabasi-Albert \cite{BA} (and earlier Simon \cite{Sim}) gave a different
 random construction where the probability of a vertex $i$ to receive
 an edge from some other vertex $j$ is not uniform and fixed, but rather depends on how many
 edges $i$  already has. This was called preferential
 attachment, that is, the chance of a vertex to receive an additional
 edge increases when  it already possesses many edges. For such a
 graph, the degree sequence behaves like $k_n \sim n^{-\kappa}$ for
 some exponent $\kappa$, typically between 2 and 3. Thus, the degree
 sequence decays like a power law, and the corresponding type of graph
 is called scale free. They also gave empirical evidence of networks
 that follow such a power law degree distribution rather than an
 exponential one. This produced a big fashion, and in its wake,
 many empirical studies appeared that demonstrated or claimed such a
 power law behavior for large classes of biological or infrastructural
 networks. Thus, scalefreeness seemed a more or less universal feature
 among graphs coming from empirical data in a wide range of
 domains. While this does provide some insights (for some systematic
 discussion, see e.g. \cite{AB,DoMe,Vaz}), for example a better
 understanding of the resilience of such graphs against random
 deletion of vertices, it also directly brings us to the two issues
 raised in the introduction:
 \begin{enumerate}
 \item A feature like scalefreeness that is essentially universal in
   empirical graphs by its very nature fails to identify what is
   specific for graphs coming from a particular domain. In other
   words, do there exist systematic structural differents for example
   between gene regulation, protein-protein interaction or neural
   networks? Or, asked differently, given an empirical graph, without
   being told where it comes from, can one identify this domain on the
   basis of certain unique qualitative features?
\item Graphs with the same degree sequence can be quite different with
  respect to other qualitative invariants like the coherence, as
  emphasized in \cite{ABJ1,ABJ2}. Also, depending on the details of the
  preferential attachment rule chosen, invariants like the average or
  maximal distance can vary widely, as observed in \cite{JJ2}. Thus, is
  there some way to encode many, or even essentially all, important
  graph invariants simultaneously in some compact manner?
 \end{enumerate}

\section{The spectrum of a graph}
In order to provide a positive answer to these question, we shall now
introduce and consider the spectrum of a (finite) graph $\Gamma$  with $N$ vertices.  For functions $v$ from the vertices of $\Gamma$ to
$\mathbb{R}$, we define the  Laplacian as
\bel{3}
\Delta v(i):=  v(i) -\frac{1}{n_i} \sum_{j, j \sim i}v(j) .
\end{equation}
(Note that in the graph theoretical literature (e.g.\cite{Bol1,GoRo,Merris,Mohar}), it is more customary
to put a factor $n_i$ in front of the right hand side in the
definition of the Laplacian. The choice of convention adopted here is
explained in \cite{J1,J2}. It is equivalent to the one in \cite{Chung}.)\\
The eigenvalue equation for $\Delta$ is
\bel{6}
\Delta u = \lambda u.
\end{equation}
A nonzero solution $u$ is called an eigenfunction for the eigenvalue
$\lambda$.
$\Delta$ then has $N$ eigenvalues, perhaps occurring with
multiplicity, that is, not necessarily all distinct. (The multiplicity
of the eigenvalue $\lambda$ is the number of linearly independent
solutions of \rf{6}.) The eigenvalues of $\Delta$ 
are real and nonnegative (because $\Delta$ is a symmetric, nonnegative
operator, see e.g. \cite{J1,J2} for details). 
The smallest eigenvalue is $\lambda =0$. Its multiplicity equals the
number of components of $\Gamma$. When $\Gamma$ is connected, that is,
has only one component (as we typically assume because one can simply
study the different components as graphs in their own right), 
this eigenvalue is simple. The other eigenvalues are positive, 
and we order the eigenvalues as\footnote{Our convention here is
  different from \cite{JJ1,J1,J2,BJ2}.}
$$ \lambda_1=0 \le \lambda_2 \le ... \le \lambda_{N}.$$
For the largest eigenvalue, we have
\bel{8}
\lambda_{N} \le 2,
\qe
with equality iff the graph is bipartite. Bipartiteness means that the
graph consists
of two disjoint classes $\Gamma', \Gamma''$ with the property that there are no
edges between vertices in the same class. Thus, a single eigenvalue
determines the global property of bipartiteness.\\
The eigenvalue $\lambda=1$ also plays a special role  as it gives some
indication of vertex or motif duplications underlying the evolution of
the graph, as systematically explored in \cite{BJ1,BJ3}. For a
biological discussion, see \cite{Ohno,Wag,WoSh}. For a general
mathematical discussion of the spectrum, see \cite{Chung,BJ2} and the
references given there. \\
What we want to emphasize here is that the spectrum constitutes an
essentially complete set of graph invariants. At least without the
qualification ``essentially'', this is not literally
true, in the following sense: there are examples of non-isomorphic
graphs with the same spectrum -- such graphs are called
isospectral (see \cite{ZW}). In other words, the spectrum of a graph does not
determine the graph completely (see \cite{IpMi} for a graph
reconstruction algorithm from the spectrum). Such isospectral graphs, while not
necessarily isomorphic, are qualitatively very similar to each other,
however. In particular, the spectrum of the graph encodes all the
essential qualitative properties of a graph, like cohesion and
coherence (e.g. \cite{Chung,JJ1,AJW}). Thus, for practical purposes, this is a good enough set of
graph invariants. \\
In conclusion, the spectrum of a graph yields a set of invariants that
on one hand captures what is specific about that graph and on the
other hand simultaneously encodes all its important properties.

\section{Visualization of a graph through its spectral plot}
\begin{figure}[h]
\begin{center}
\includegraphics[width=.5\textwidth]{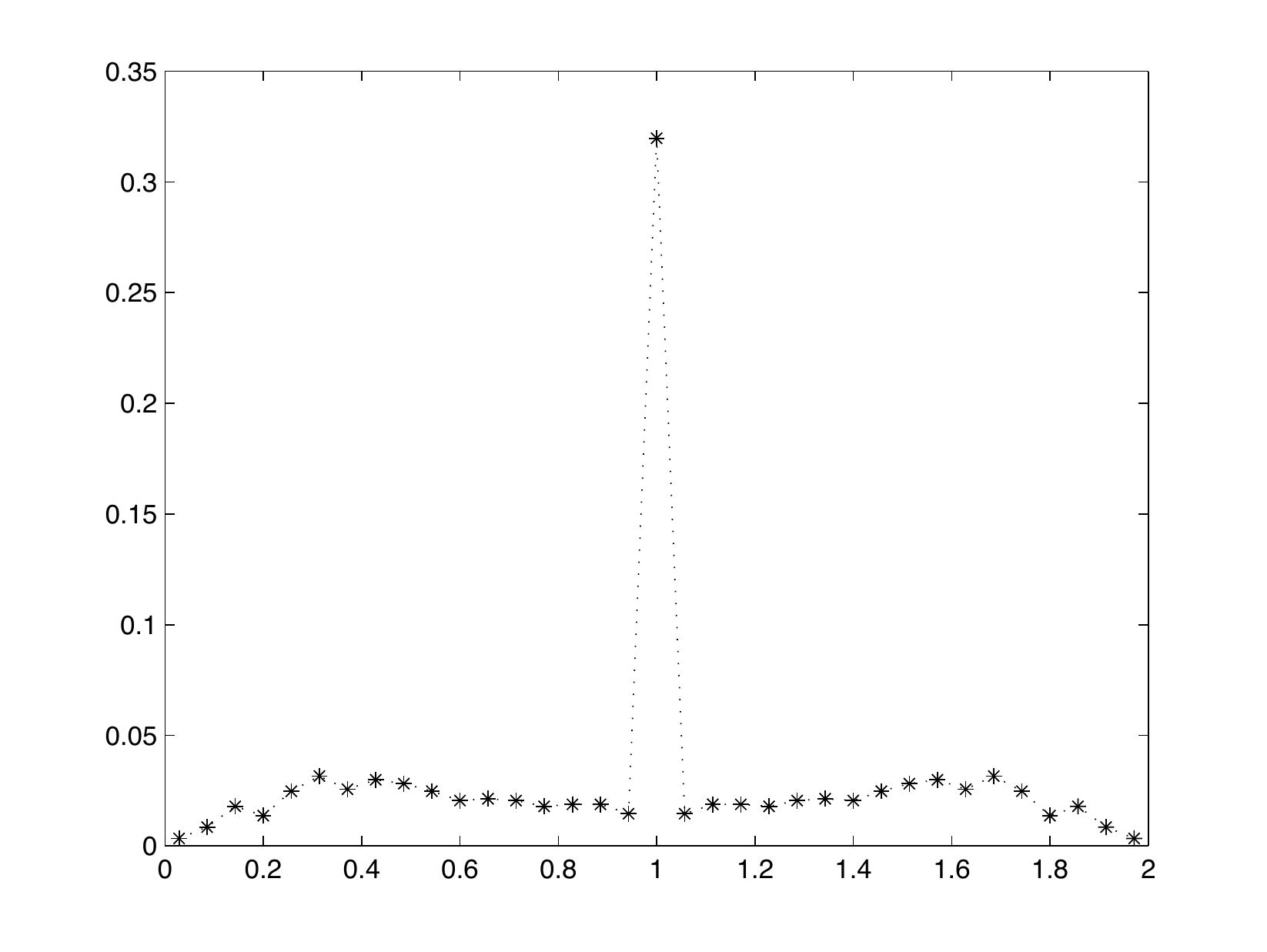}\includegraphics[width=.5\textwidth]{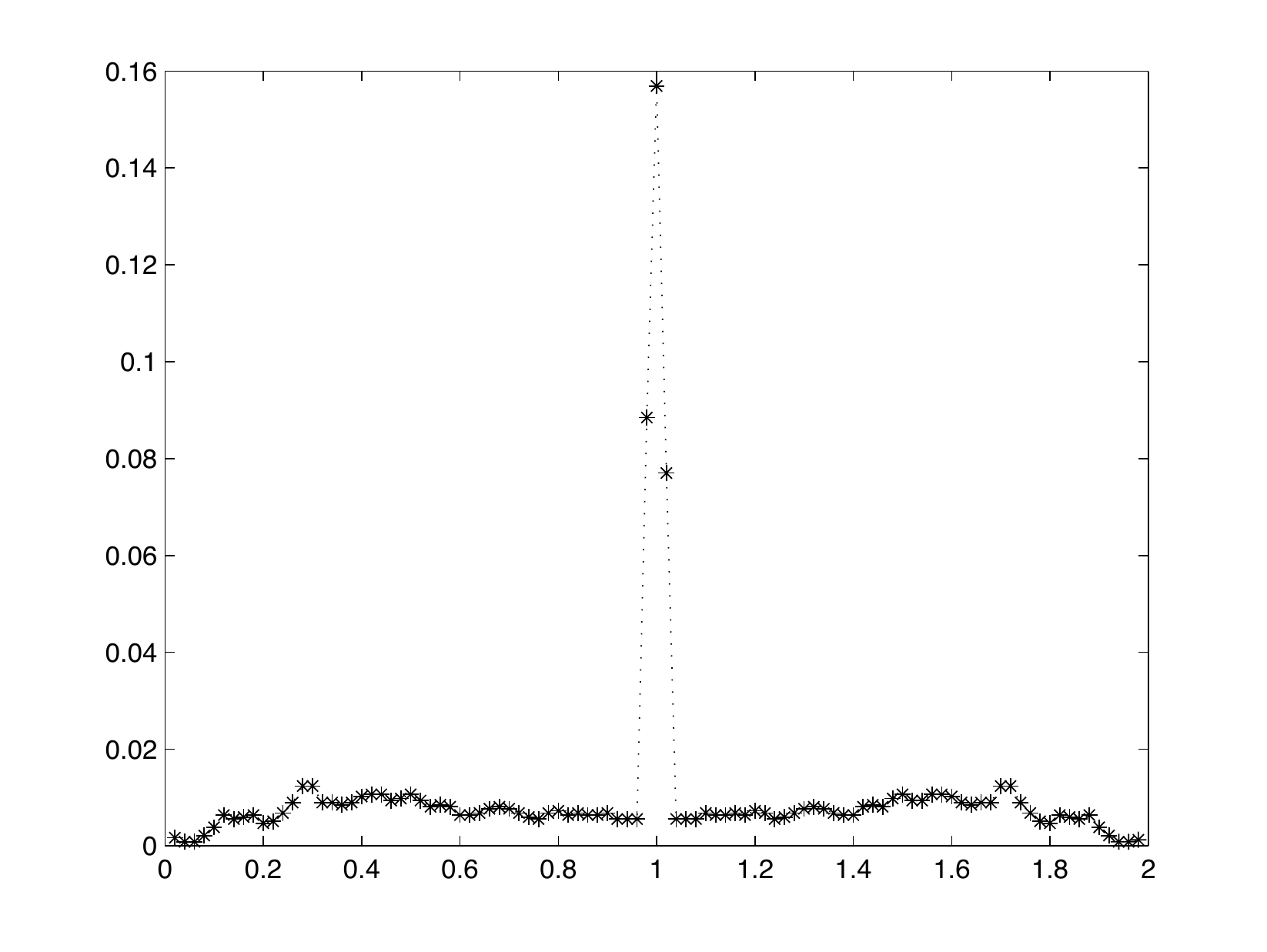}\\

(a)\hspace{.5\textwidth}(b)\\

\includegraphics[width=.5\textwidth]{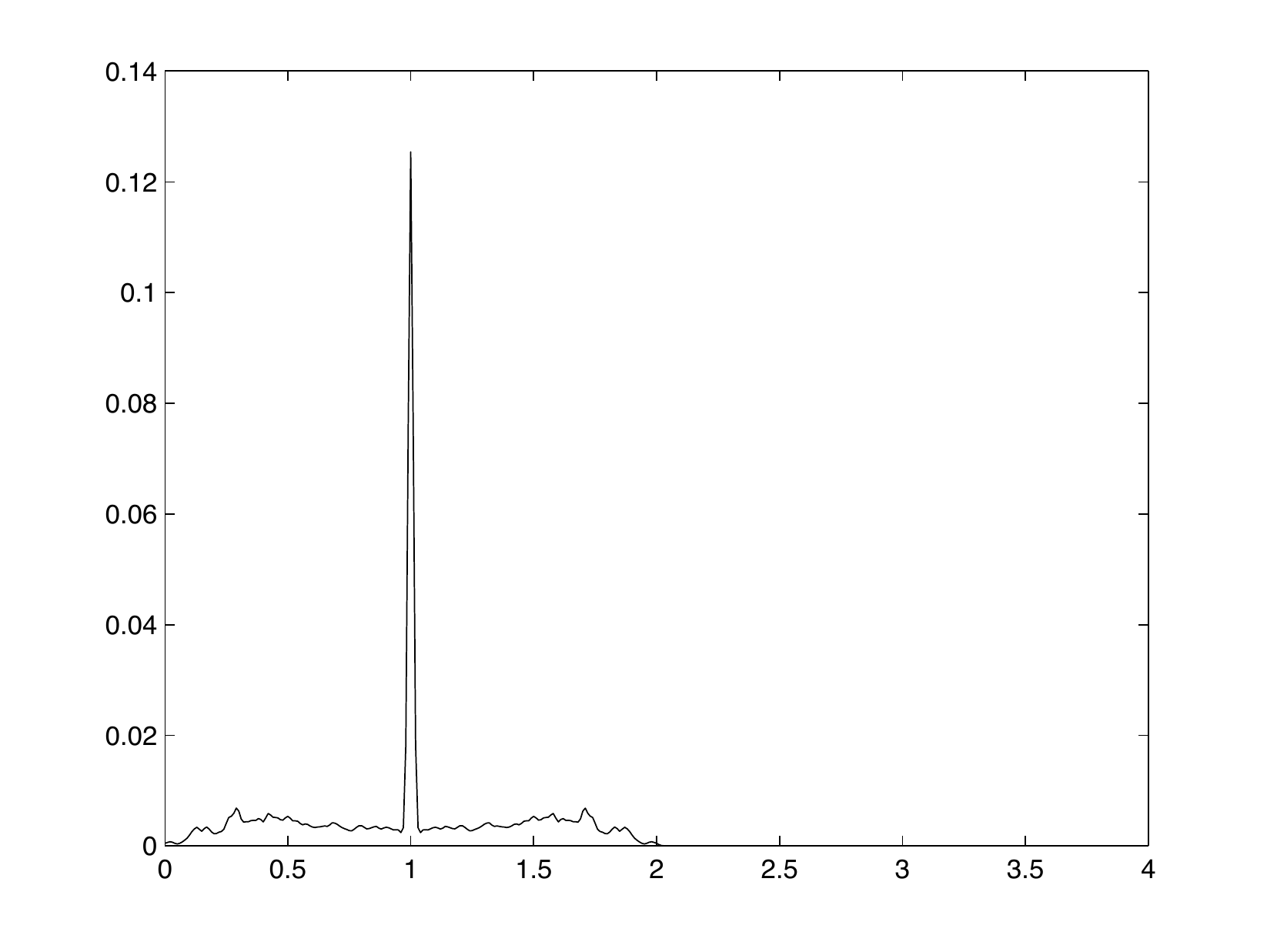}\includegraphics[width=.5\textwidth]{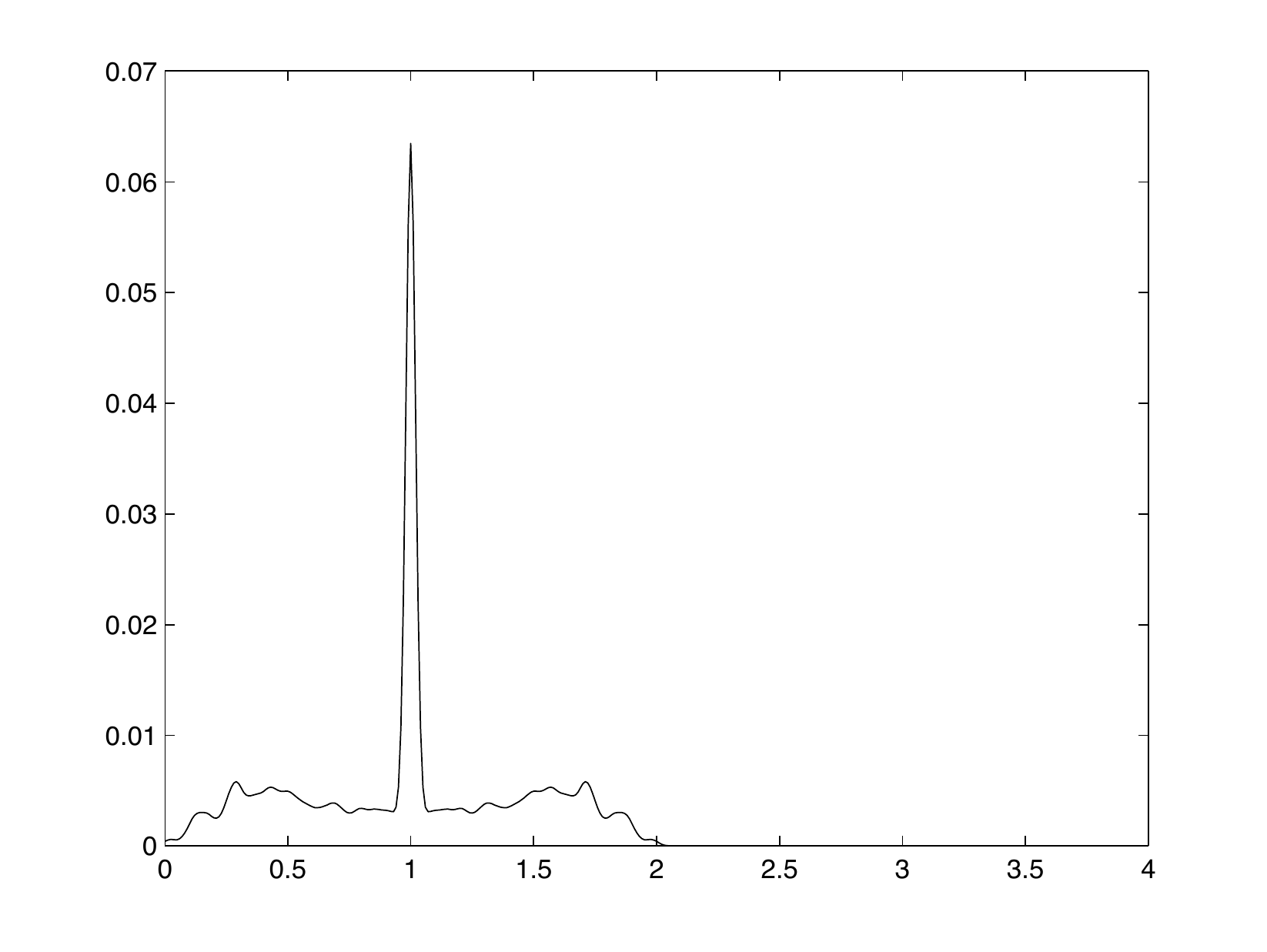}\\

(c)\hspace{.5\textwidth}(d)\\

\includegraphics[width=.5\textwidth]{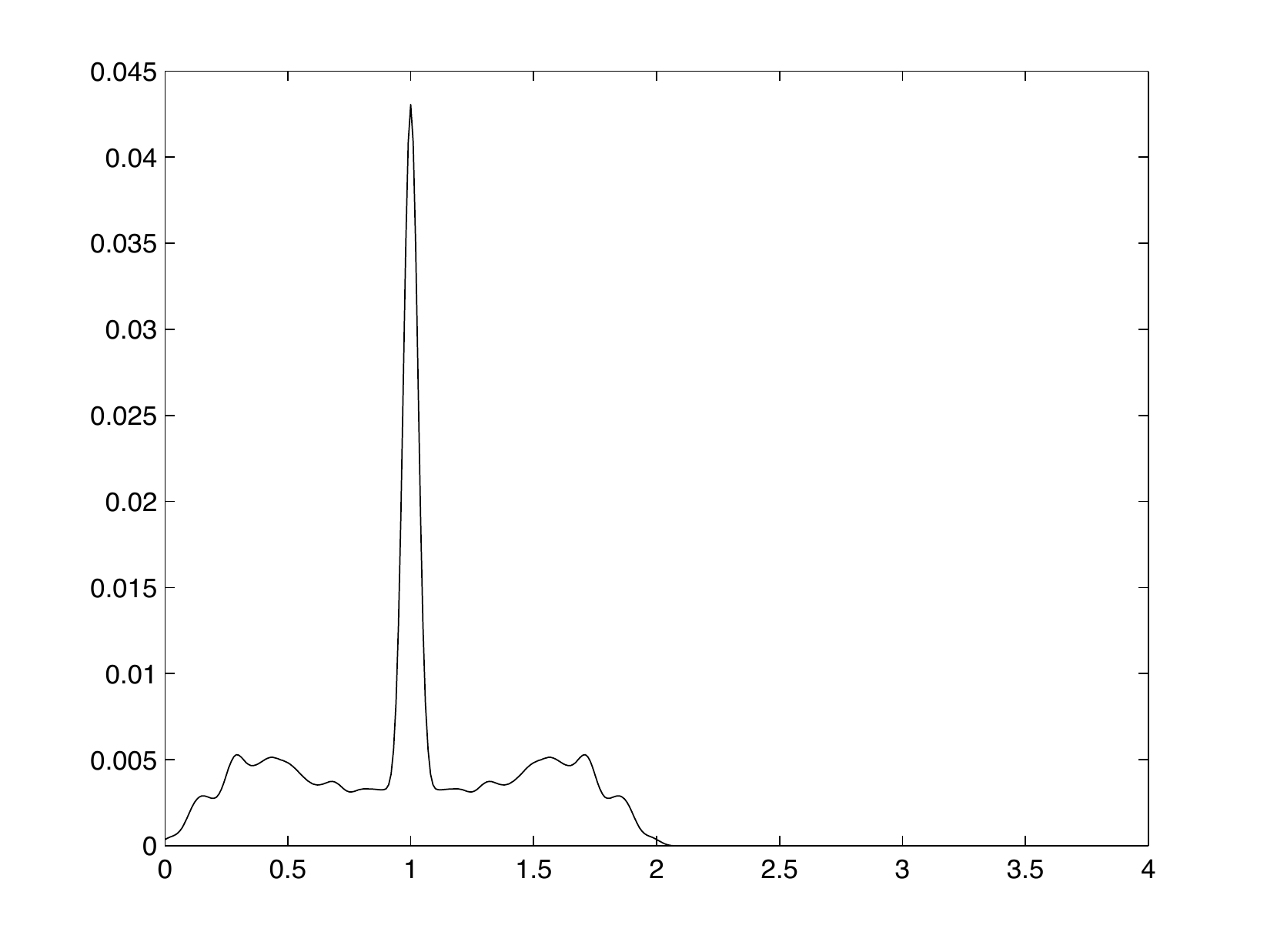}\includegraphics[width=.5\textwidth]{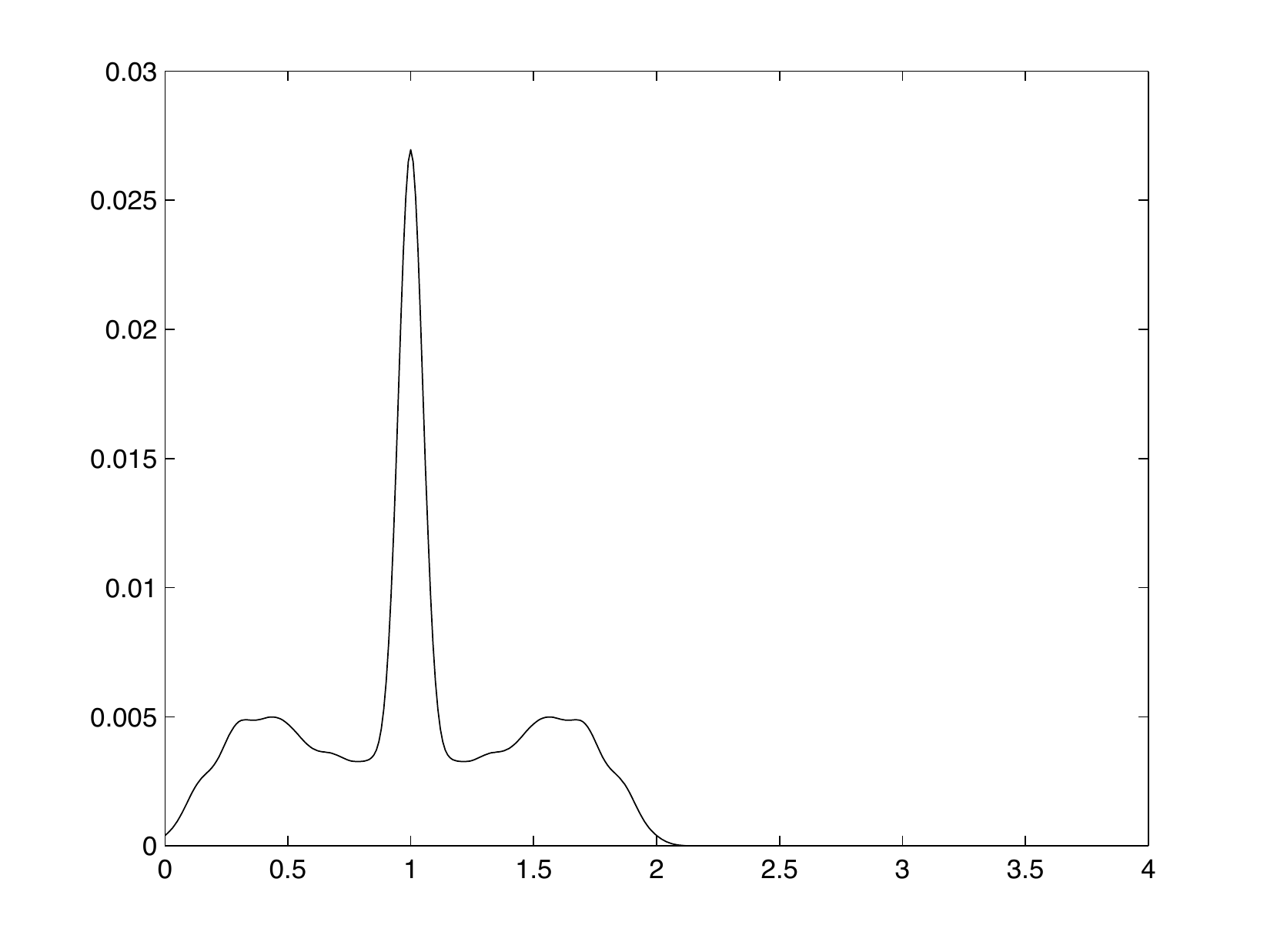}\\

(e)\hspace{.5\textwidth}(f)\\

\end{center}
\caption{Spectral plots of metabolic network of {\it Caenorhabditis elegans}. Size of the network is 1173. Nodes are substrates, enzymes and intermediate complexes. Data used in \cite{JeongEtAl2000}. Data Source: http://www.nd.edu/$\sim$networks/resources.htm/. [Download date: 22nd Nov. 2004]. (a) Relative frequency plot with number of bins = 35. (b) Relative frequency polygon with overlapping bins with bin width = 0.04 and number of bins = 99 and bins are taken as $[0, .04], [.02, .06], [.04, .08], \dots , [1.96, 2]$. (c) with Gaussian kernel with $\sigma= 0.01$. (d) with Gaussian kernel with $\sigma= 0.02$. (e) with Gaussian kernel with $\sigma= 0.03$. (f) with Gaussian kernel with $\sigma= 0.05$.}
\label{MetabolicFigExmp}
\end{figure}

\begin{figure}[h]
\begin{center}
\includegraphics[width=.5\textwidth]{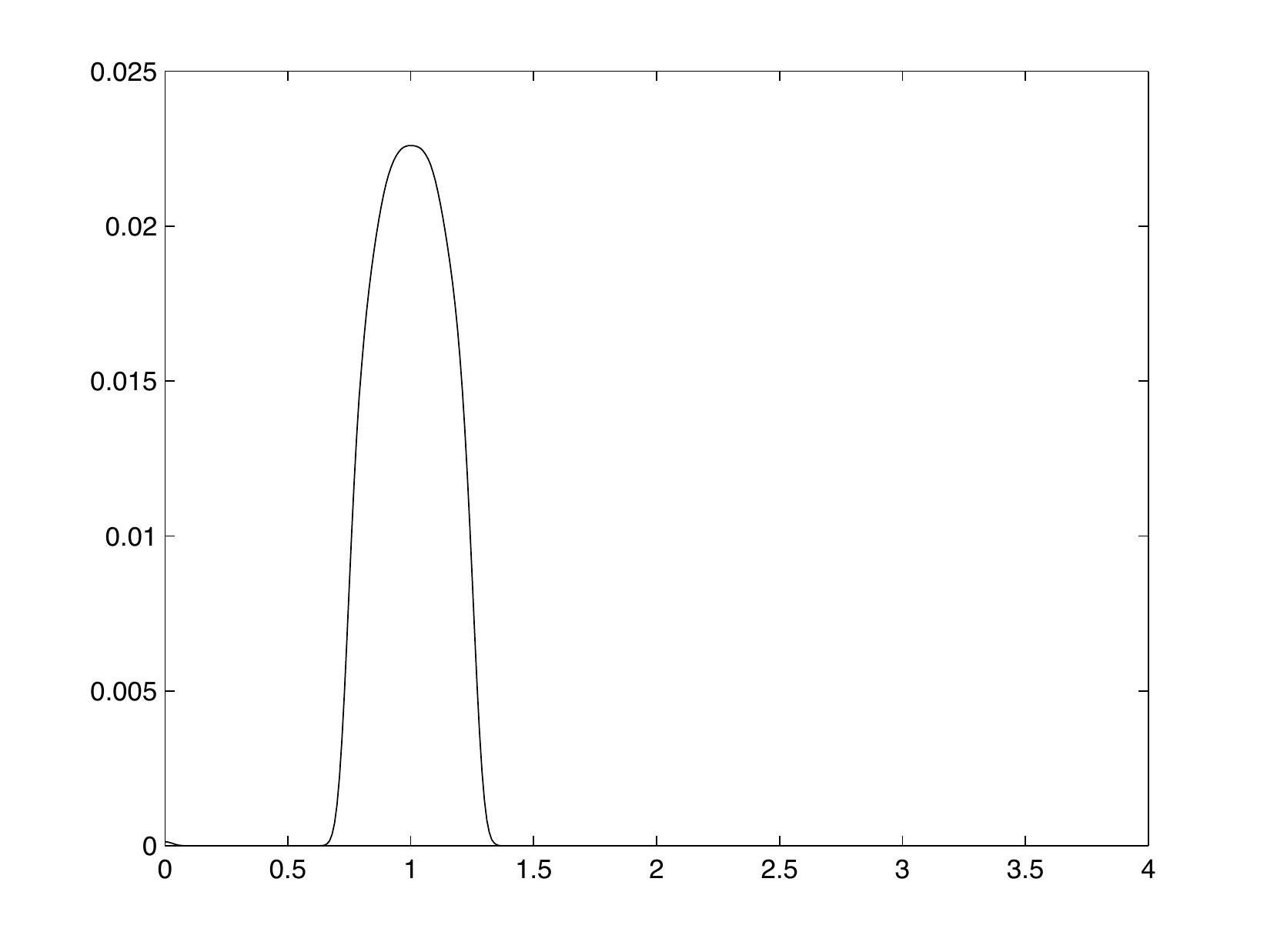}\includegraphics[width=.5\textwidth]{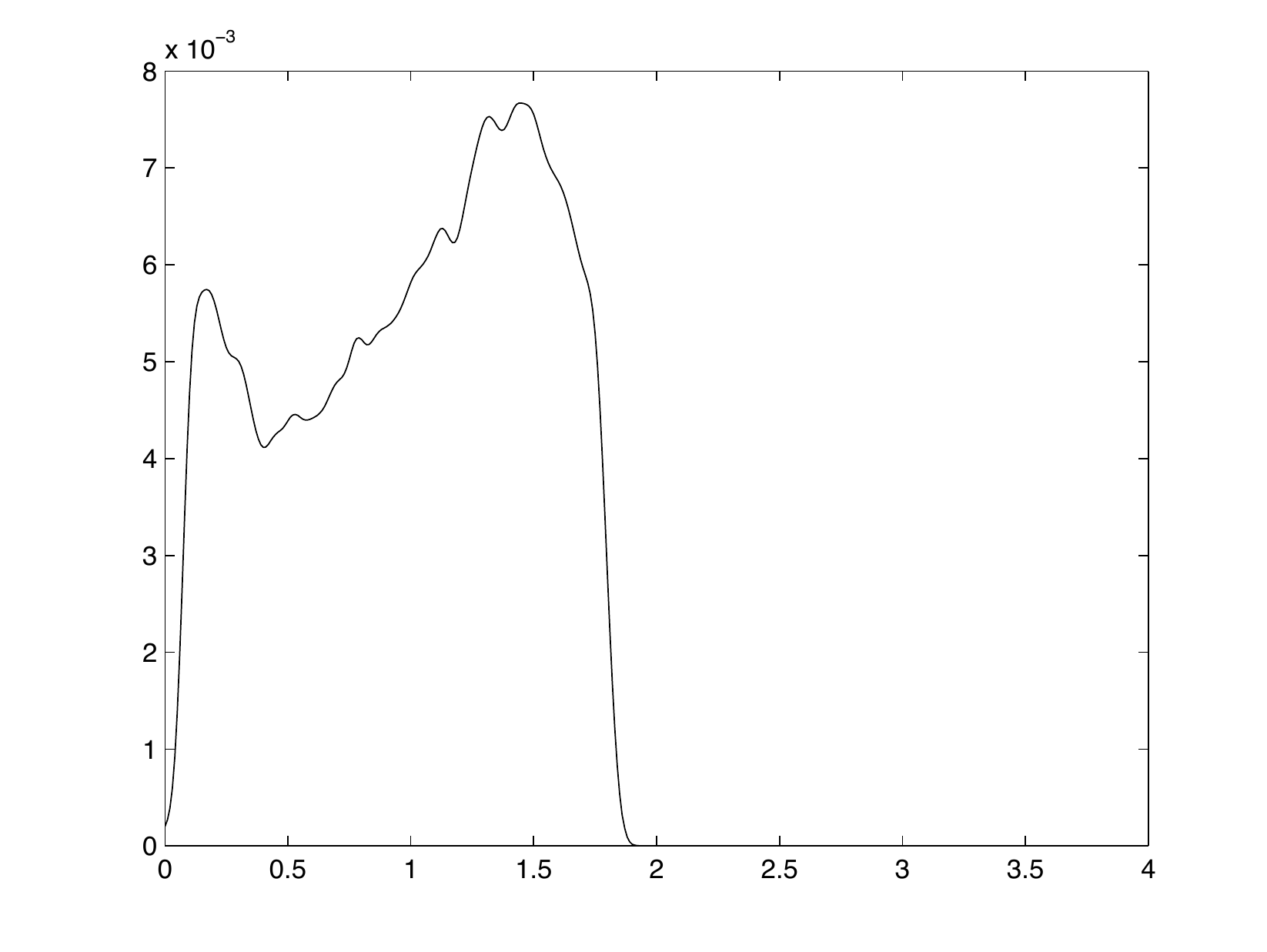}\\
(a)\hspace{.5\textwidth}(b)\\
\includegraphics[width=.5\textwidth]{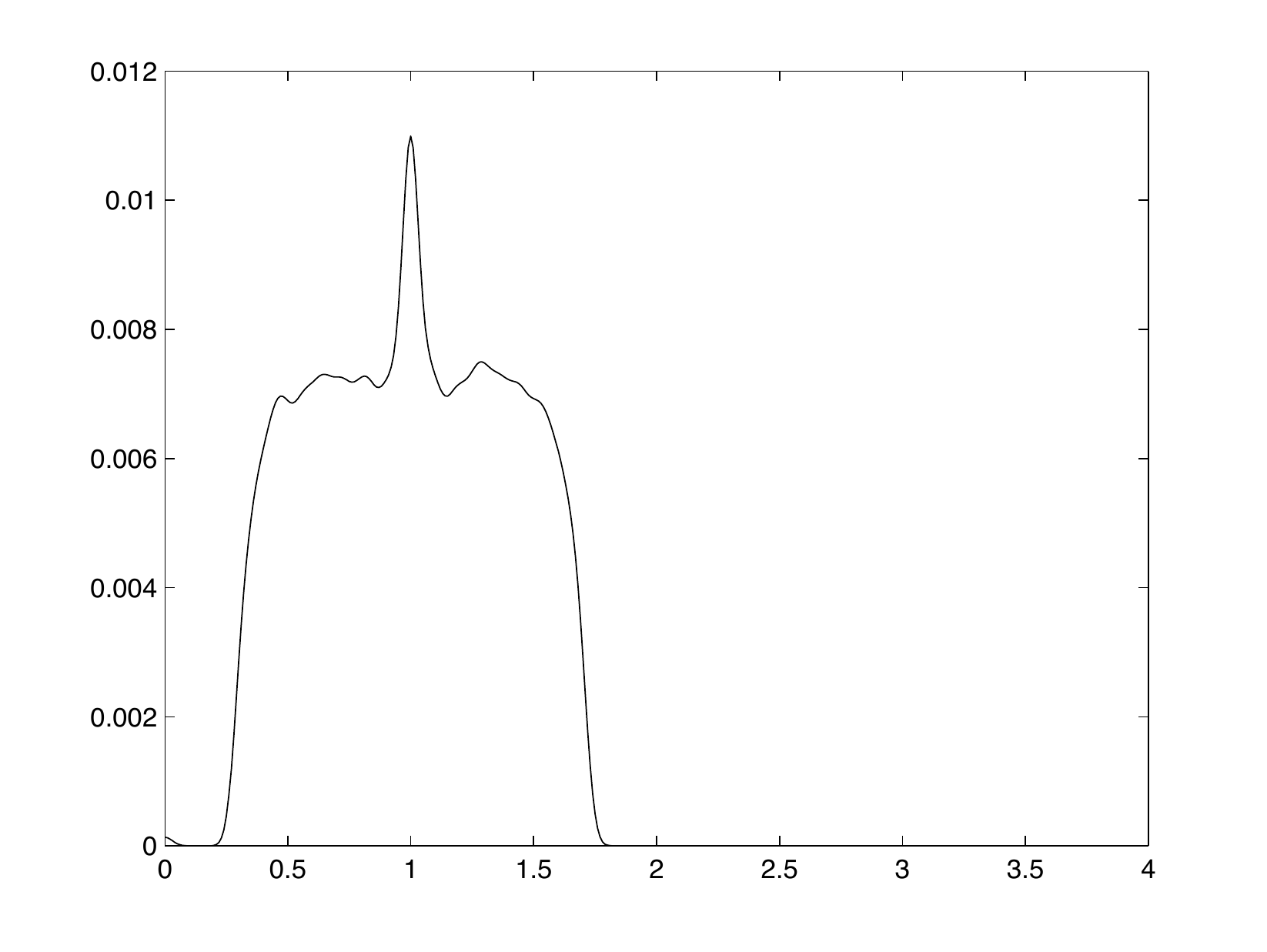}\\
(c)

\end{center}
\caption{Specral plots of generic networks. (a) Random network by Erd\"{o}s and R\'{e}nyi's model \cite{ErdosRenyi1959} with $p = 0.05$. (b) Small-world network by Watts and Stogatz model \cite{WattsStrogatz1998} (rewiring a regular ring lattice of average degree 4 with rewiring probability $0.3$). (d) Scale-free network by Albert and Barab\'{a}si model \cite{BA} ($m_0 = 5 \text{ and } m = 3 $). {\it Size of all networks is $1000$. All figures are ploted with $100$ realization}.}
\label{GeneralModelsFigs}
\end{figure}

\begin{figure}[h]
\begin{center}
\includegraphics[width=.5\textwidth]{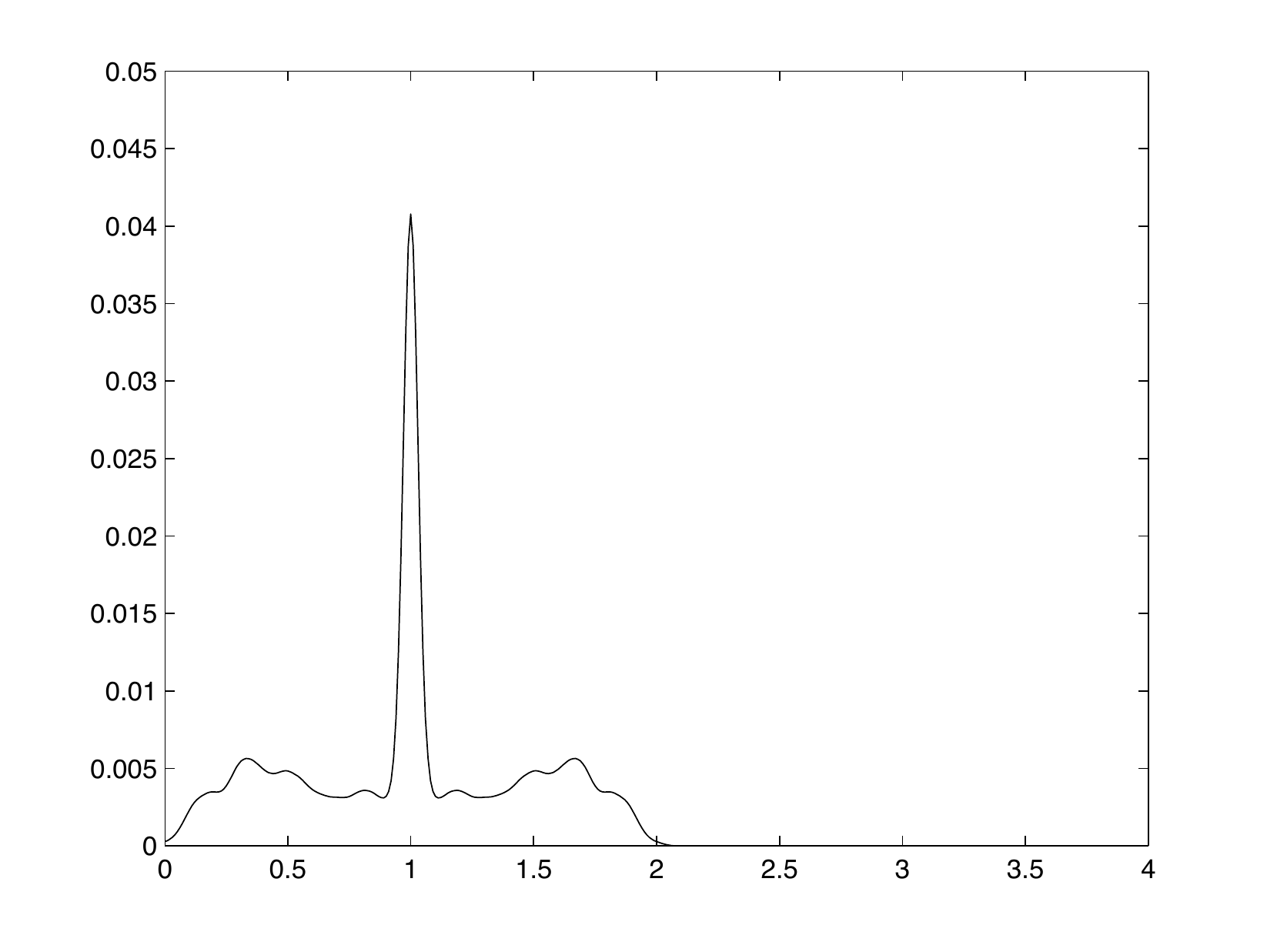}\includegraphics[width=.5\textwidth]{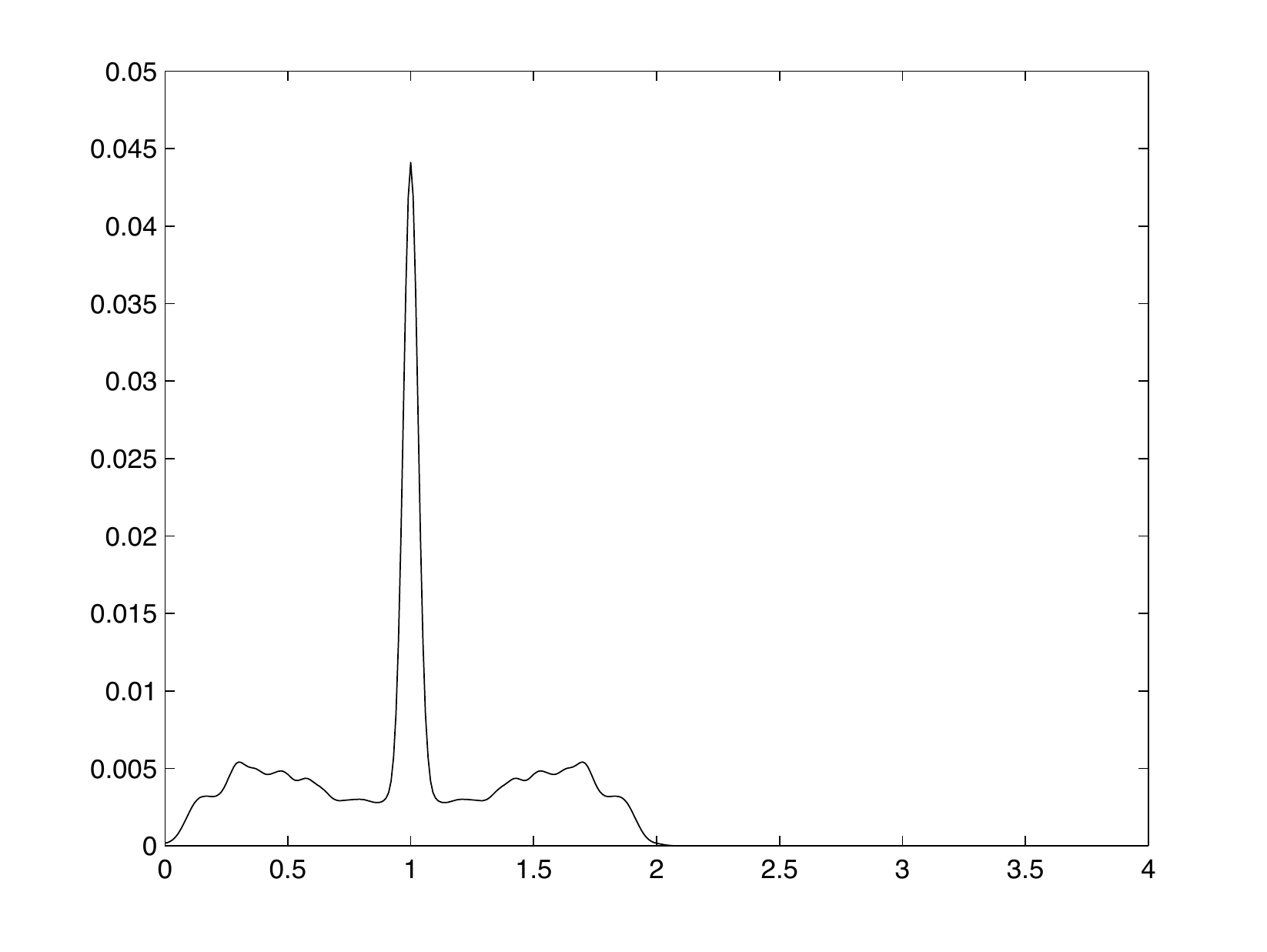}\\
(a)\hspace{.5\textwidth}(b)\\
\includegraphics[width=.5\textwidth]{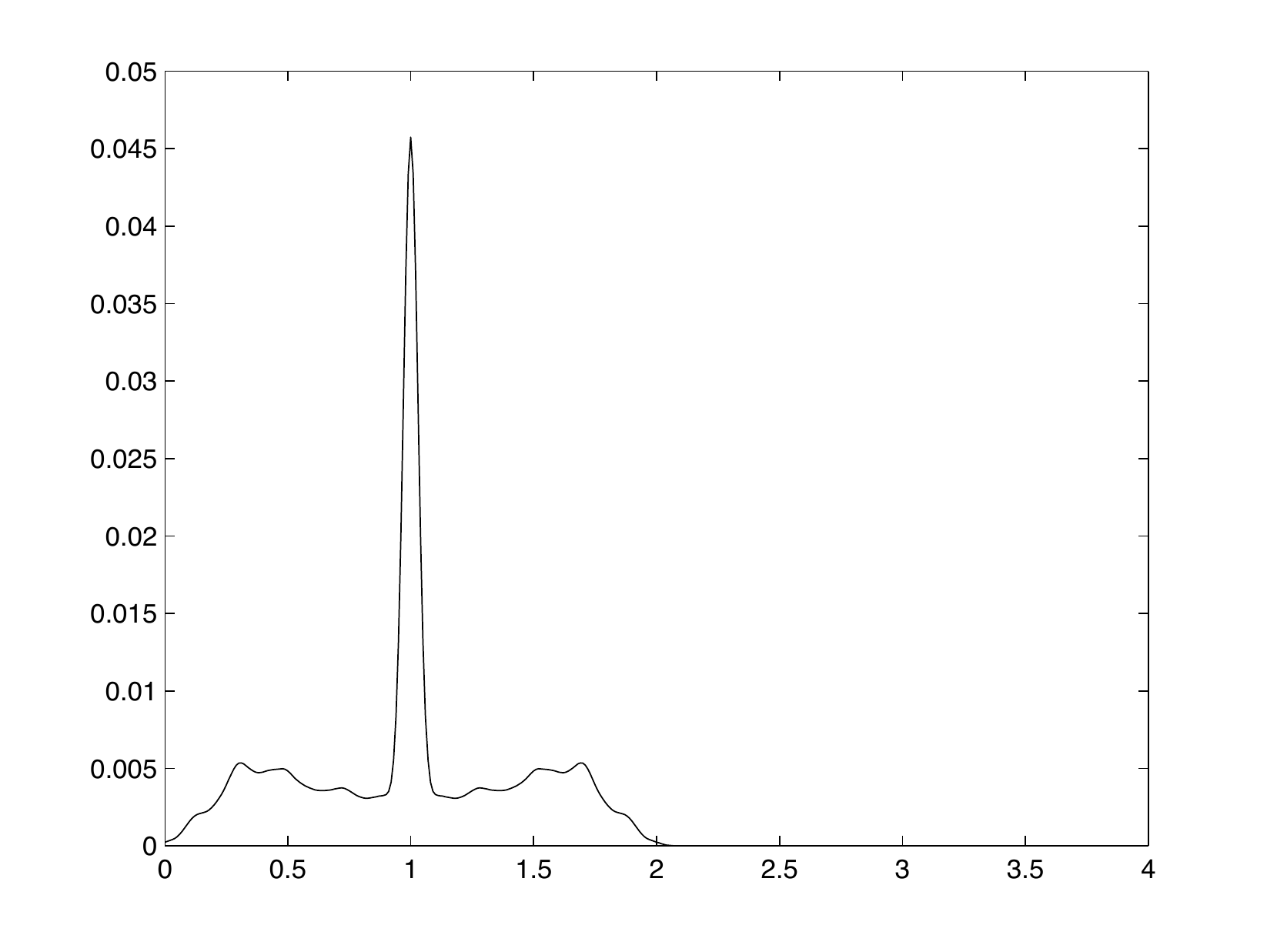}\\
(c)
\end{center}
\caption{Metabolic networks. Here nodes are substrates, enzymes and intermediate complexes. Data used in \cite{JeongEtAl2000}. Data Source: http://www.nd.edu/$\sim$networks/resources.htm/. [Download date: 22nd Nov. 2004]  (a) {\it Pyrococcus furiosus}. Network size is 746. (b) {\it Aquifex aeolicus}. Network size is 1052. (c) {\it Saccharomyces cerevisiae}. Network size is 1511.}
\label{MetabolicFigs}
\end{figure}

\begin{figure}[h]
\begin{center}
\includegraphics[width=.5\textwidth]{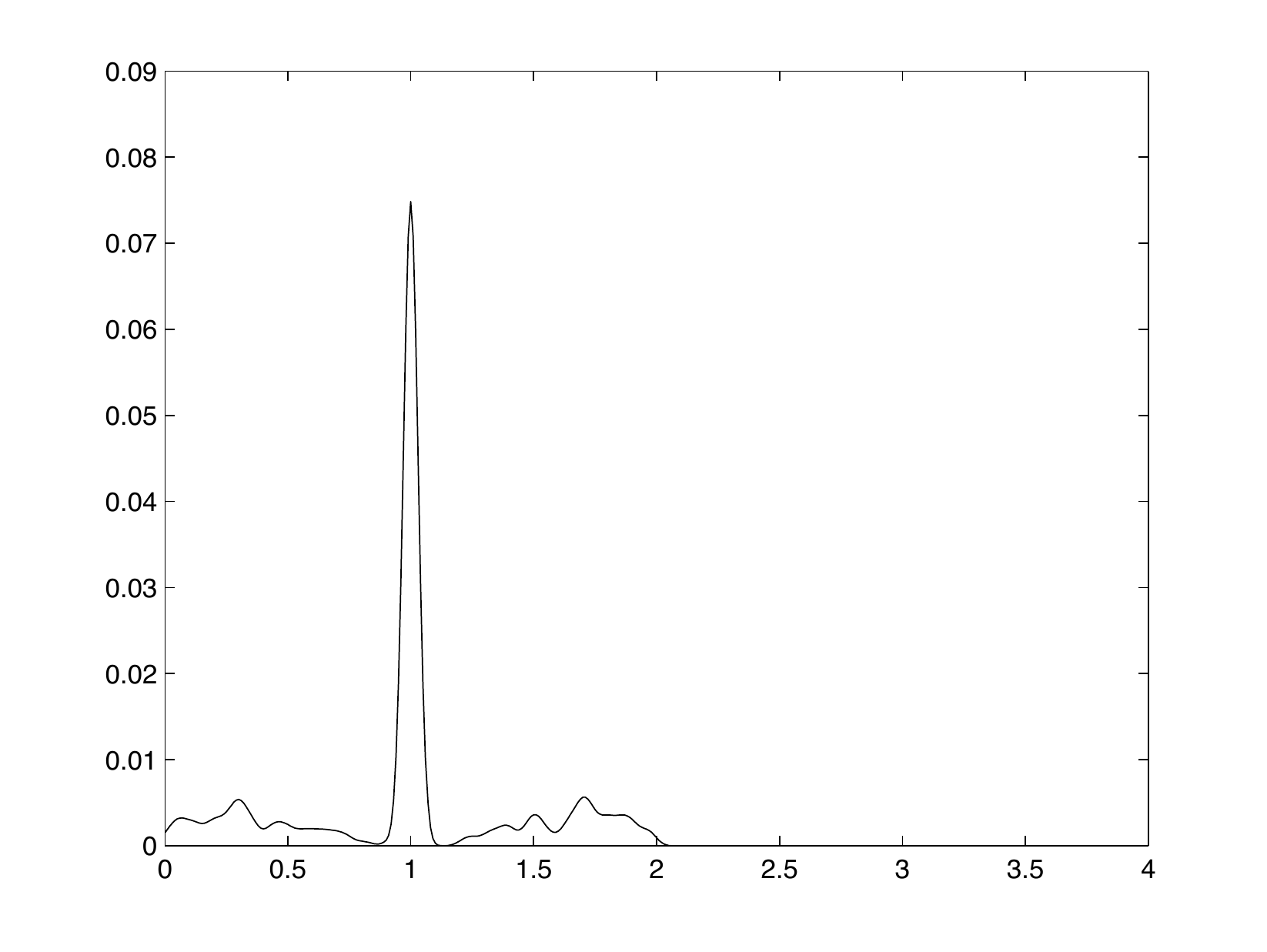}\includegraphics[width=.5\textwidth]{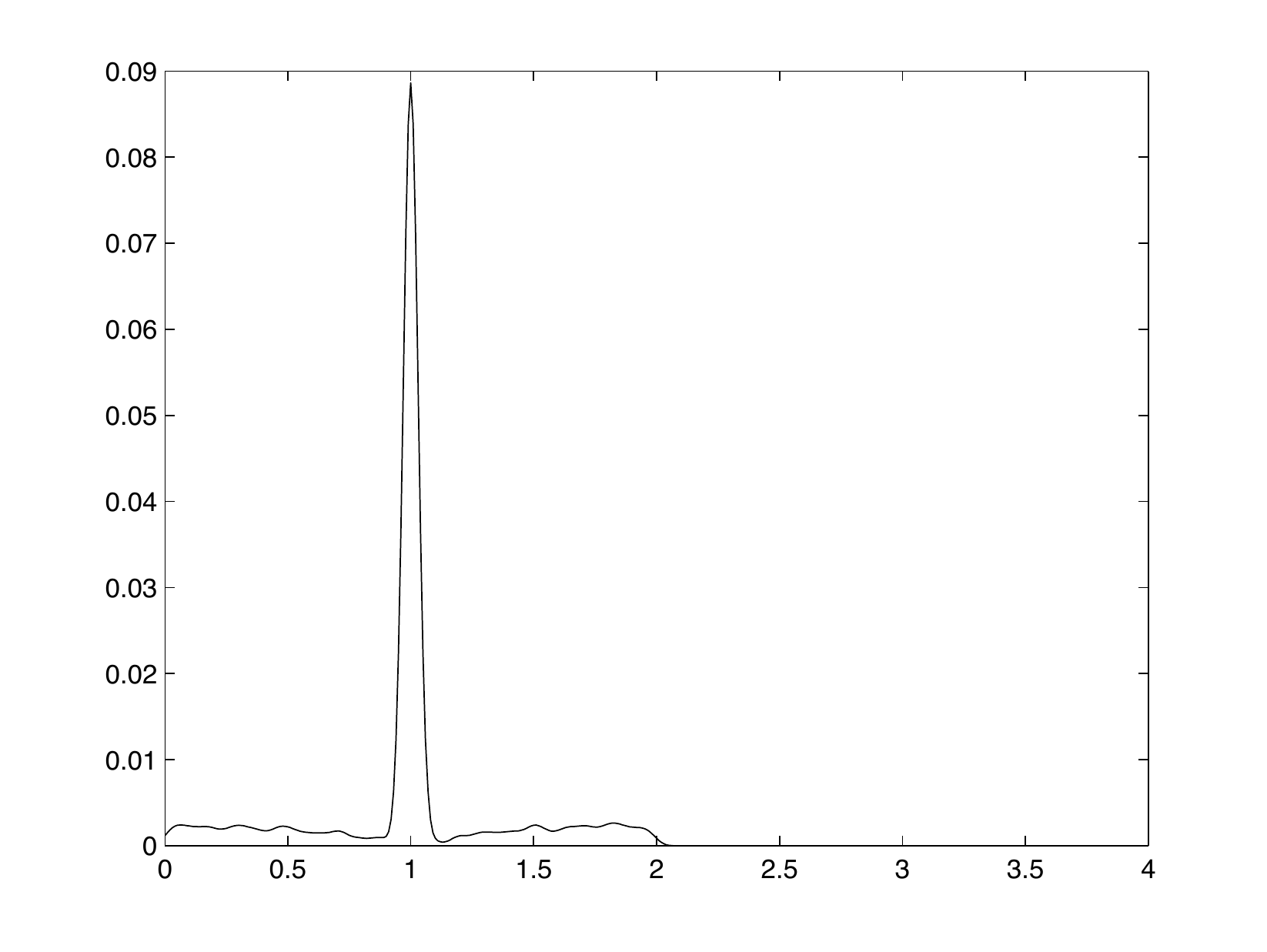}\\

(a)\hspace{.5\textwidth}(b)\\

\end{center}
\caption{Transcription networks. Data source: Data published by Uri Alon (http://www.weizmann.ac.il/mcb/UriAlon/ ). [Download date: 13th Oct. 2004]. Data used in \cite{MiloEtAl2002,Shen-OrrEtAl2002}. (a) {\it Escherichia coli}.  Size of the network is 328. (b) {\it Saccharomyces cerevisiae}. Size of the network is 662.}
\label{TranscripFigs}
\end{figure}

\begin{figure}[h]
\begin{center}
\includegraphics[width=.5\textwidth]{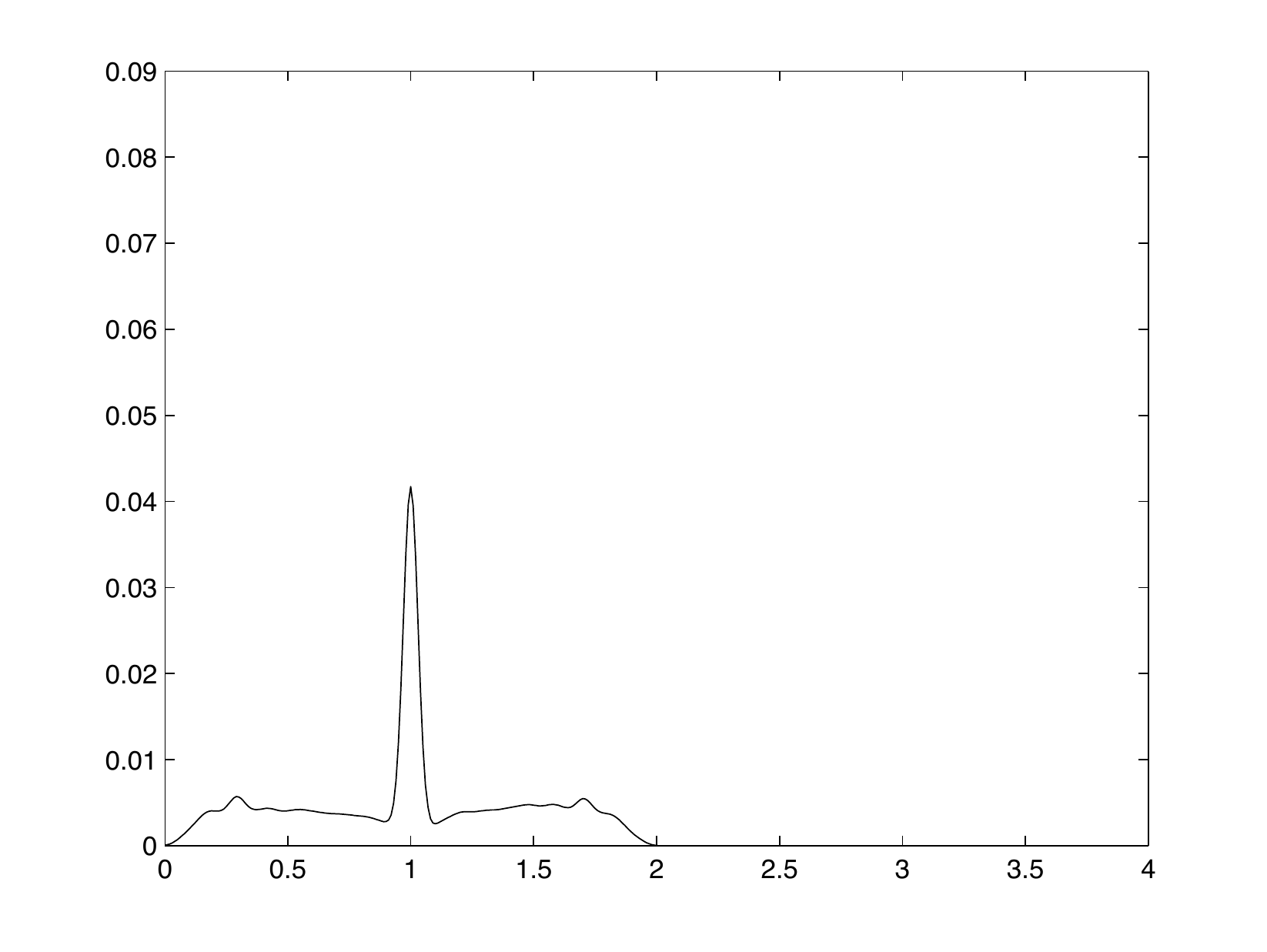}\includegraphics[width=.5\textwidth]{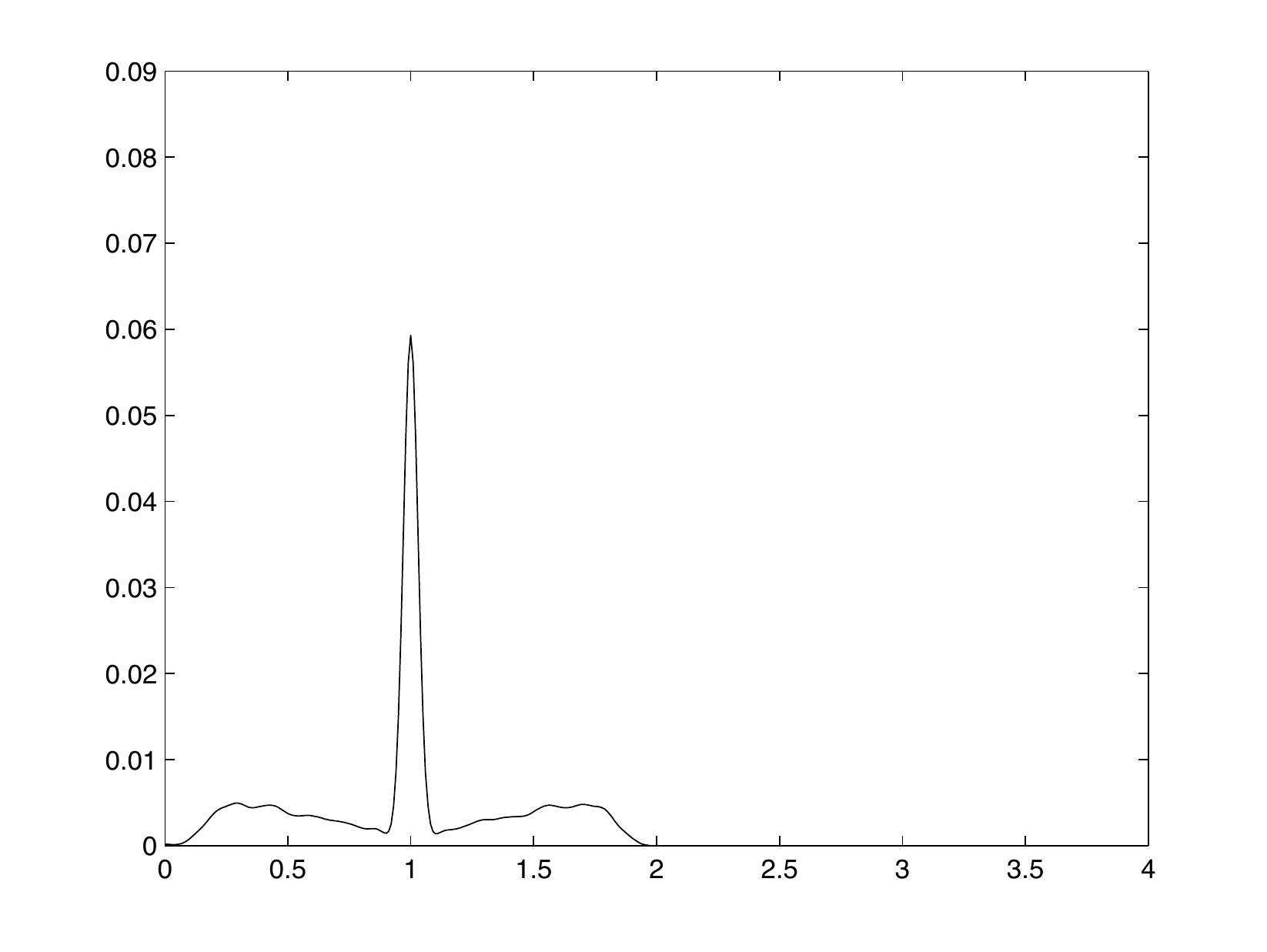}\\
(a)\hspace{.5\textwidth}(b)\\
\includegraphics[width=.5\textwidth]{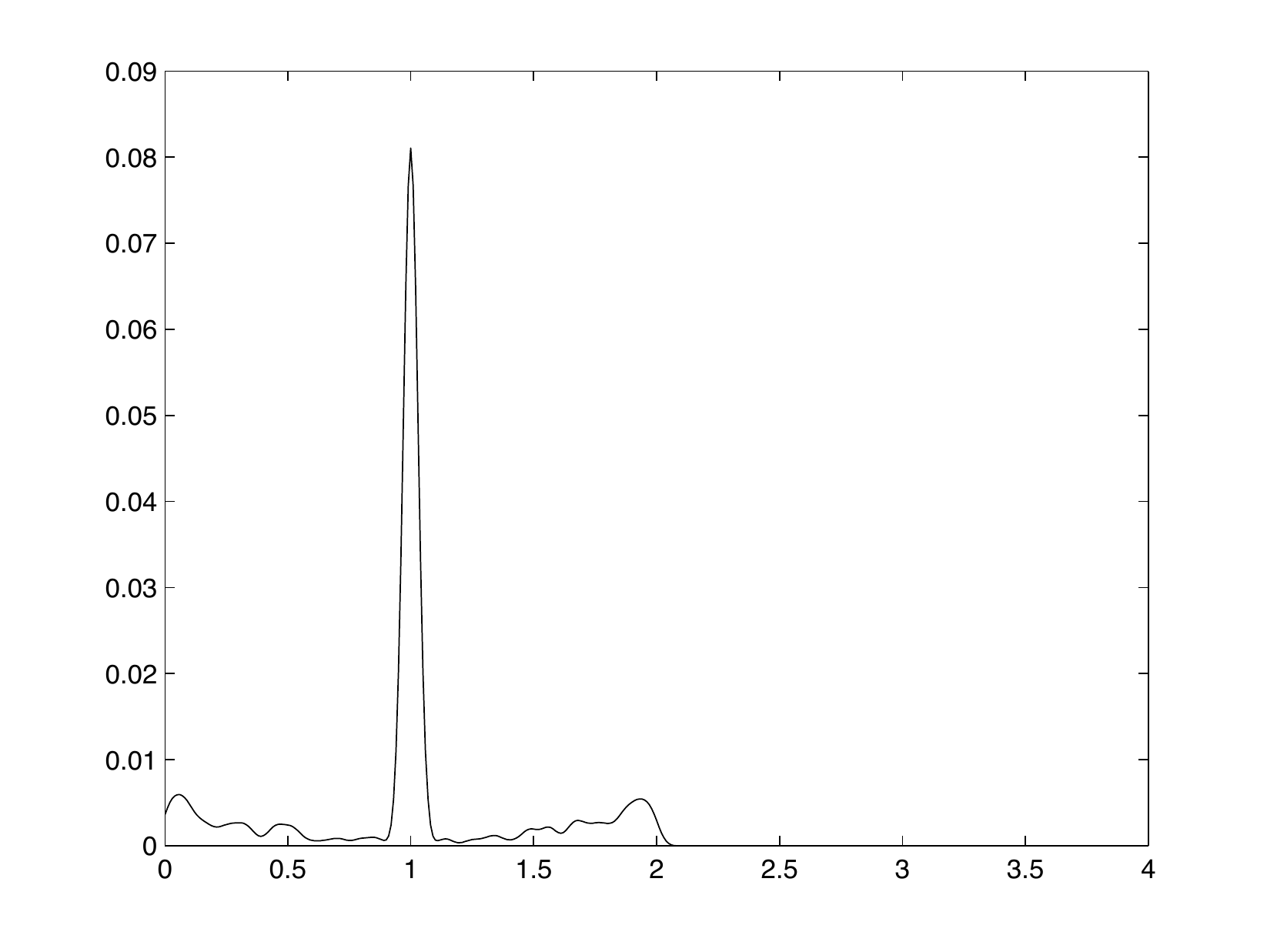}\\
(c)
\end{center}
\caption{Protein-protein interaction networks. Data are collected from http://www.cosin.org/ [download date: 25th September, 2005]. (a) {\it Saccharomyces cerevisiae}. Size of the network is 3930. (b) {\it Helicobacter pylori}. Size of the network is 710. (c) {\it Caenorhabditis elegans}. Size of the network is 314.}
\label{PPINFigs}
\end{figure}

\begin{figure}[h]
\begin{center}
\includegraphics[width=.5\textwidth]{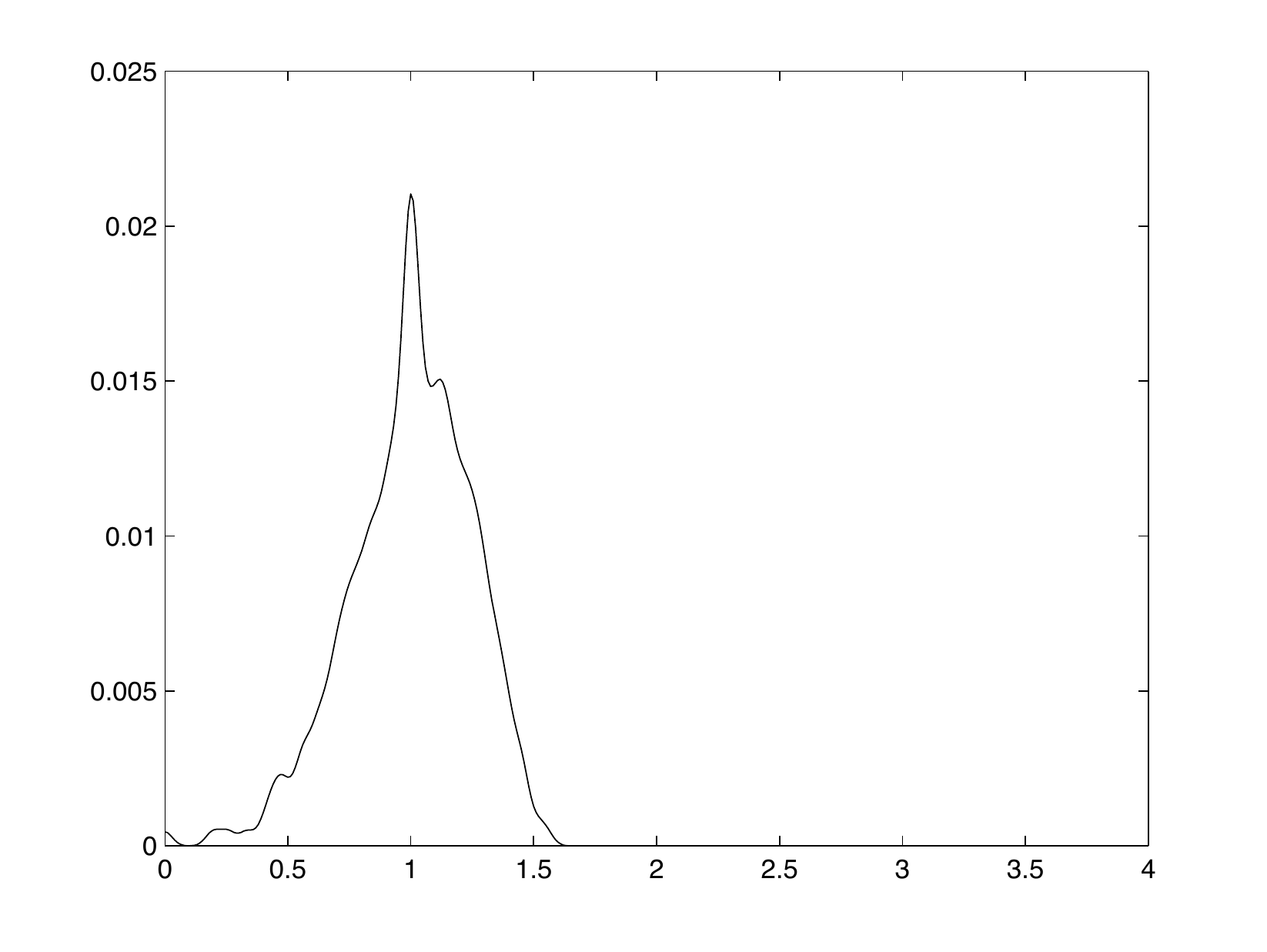}\includegraphics[width=.5\textwidth]{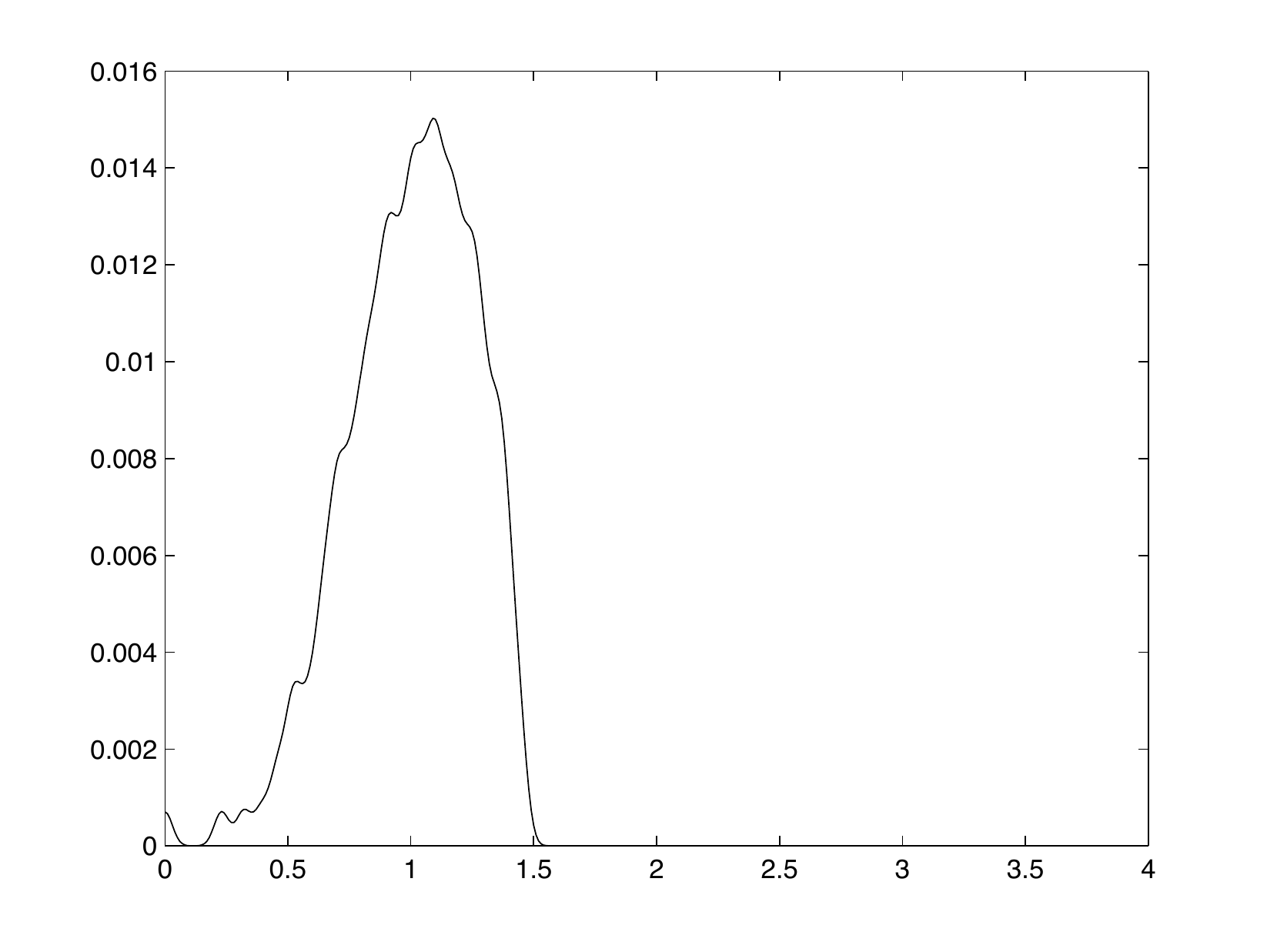}\\
(a)\hspace{.5\textwidth}(b)\\
\includegraphics[width=.5\textwidth]{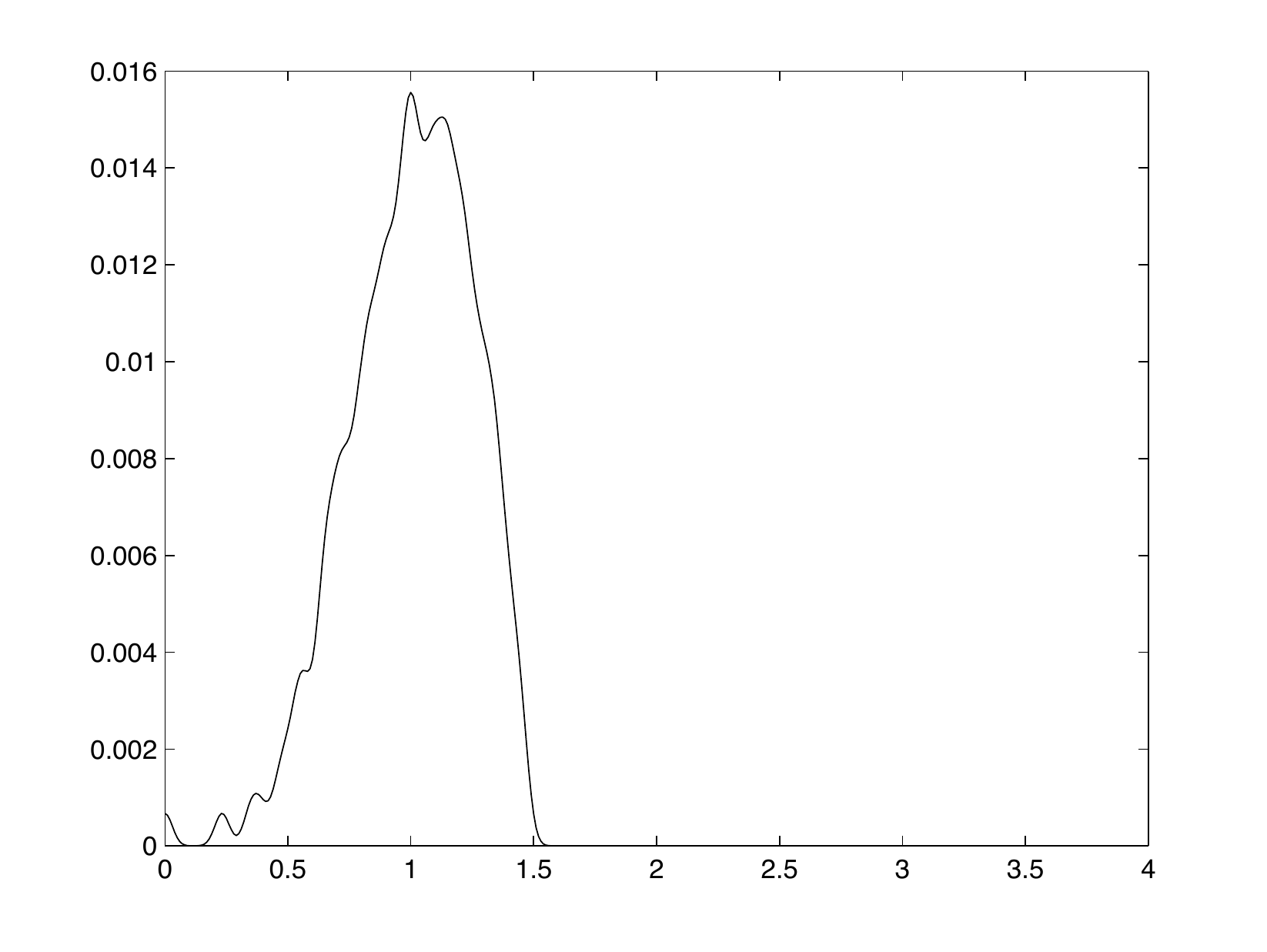}\\
(c)
\end{center}
\caption{Neuronal connectivity. (a) {\it Caenorhabditis elegans}. Size of the network is 297. Data used in \cite{WattsStrogatz1998,WhiteEtAl1986}. Data Source: http://cdg.columbia.edu/cdg/datasets/ [Download date: 18th Dec. 2006]. (b) {\it Caenorhabditis elegans} (animal JSH, L4 male) in the nerve ring and RVG regions. Network size is 190. Data source: Data is assembled by J. G. White, E. Southgate, J. N. Thomson, S. Brenner \cite{WhiteEtAl1986} and was later revisited by R. M. Durbin (Ref. http://elegans.swmed.edu/parts/ ). [Download date: 27th Sep. 2005]. (c) {\it Caenorhabditis elegans} (animal  N2U, adult hermaphrodite) in the nerve ring and RVG regions. Network size is 199. Data source: Data is assembled by J. G. White, E. Southgate, J. N. Thomson, S. Brenner \cite{WhiteEtAl1986} and was later revisited by R. M. Durbin (Ref. http://elegans.swmed.edu/parts/ ). [Download date: 27th Sep. 2005].}
\label{NeuroFigs}
\end{figure}

\begin{figure}[h]
\begin{center}
\includegraphics[width=.5\textwidth]{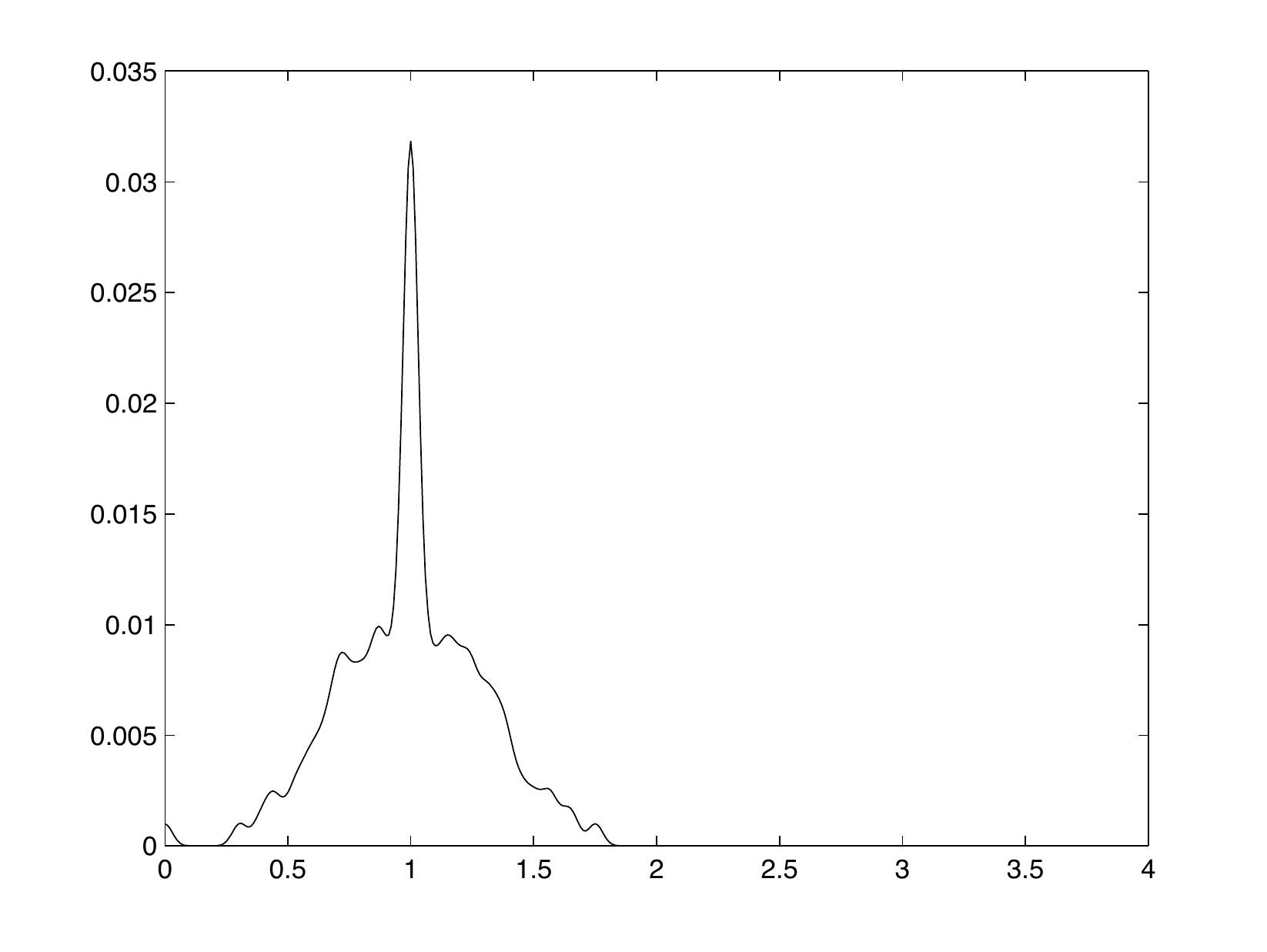}\includegraphics[width=.5\textwidth]{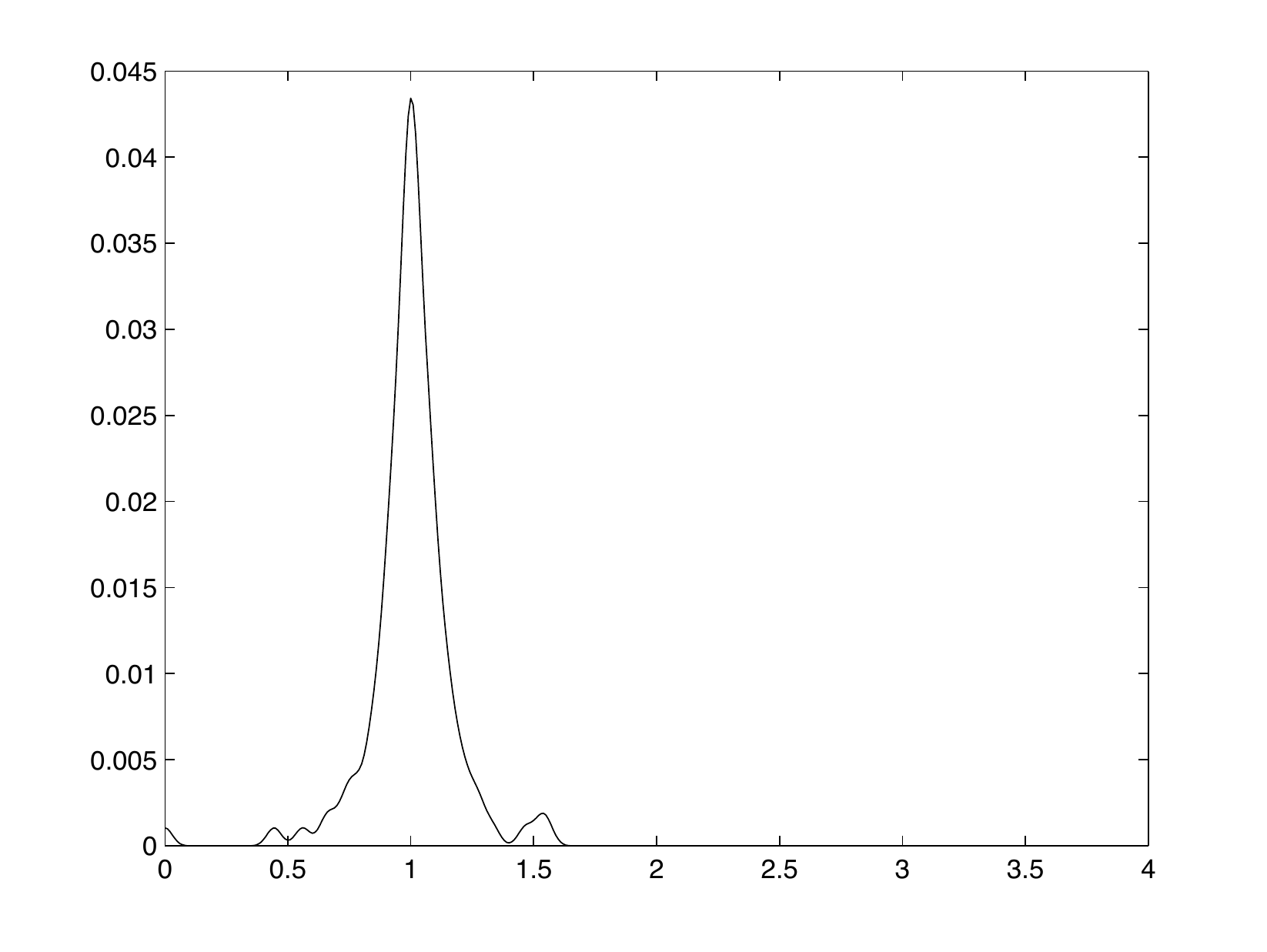}\\
(a)\hspace{.3\textwidth}(b)\\
\includegraphics[width=.5\textwidth]{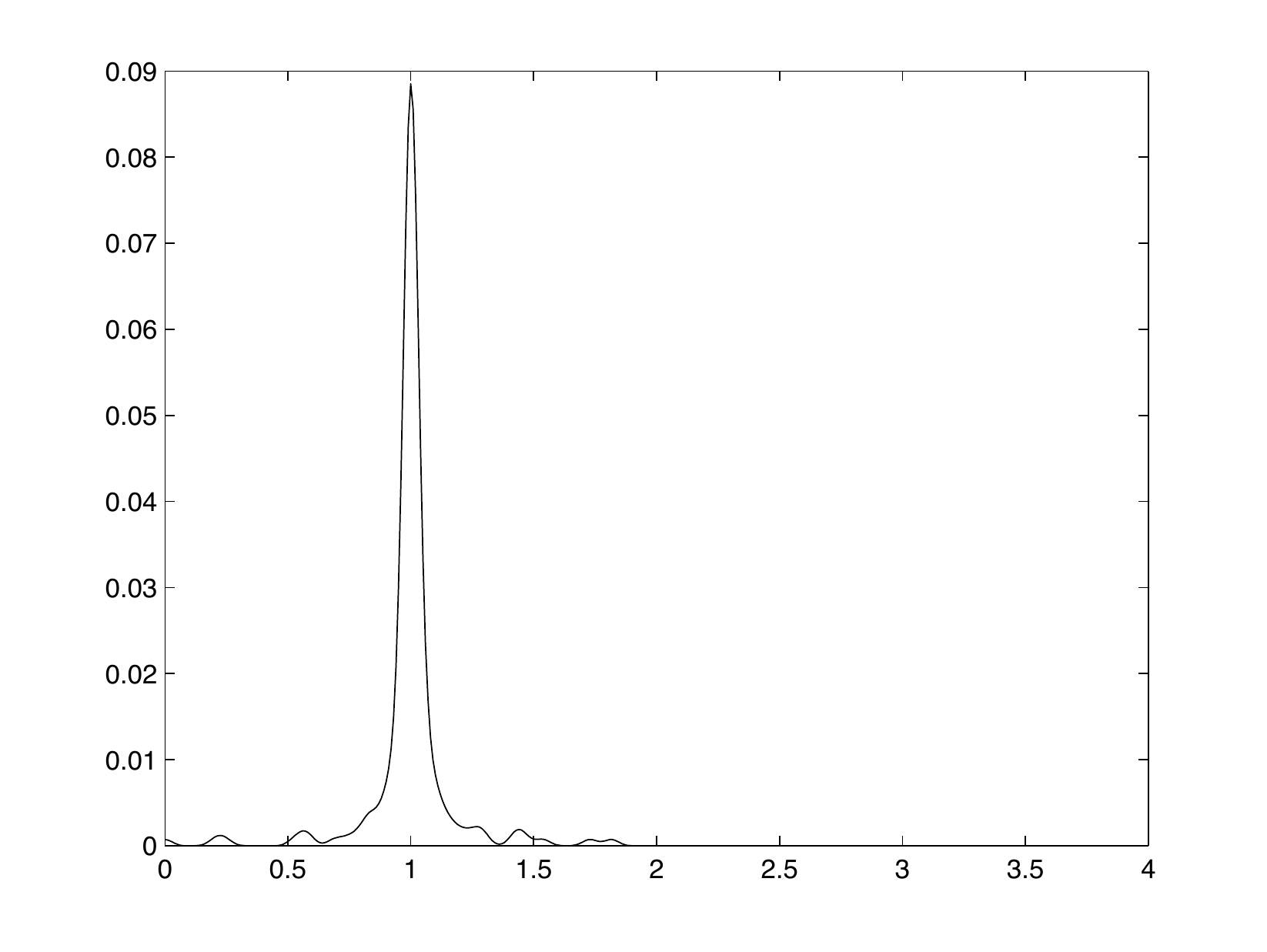}\\
(c)
\end{center}
\caption{Food-web. (a) From "Ythan estuary". Data downloaded from http://www.cosin.org/. [Download Date 21st December, 2006]. Size of the network is 135. (b) From "Florida bay in dry season". Data downloaded from http://vlado.fmf.uni-lj.si/pub/networks/data/ (main data resource: Chesapeake Biological Laboratory. Web link: http://www.cbl.umces.edu/). [Download Date 21st December, 2006]. Size of the network is 128. (c) From "Little rock lake". Data downloaded from http://www.cosin.org/. [Download Date 21st December, 2006]. Size of the network is 183.}
\label{FoodWebFigs}
\end{figure}

What is more, the spectral plot of a graph is much better amenable to
visual inspection than a direct plot of the graph or any other method
of representation that we know of. In other words, with a little
experience in graph theory, one can quickly detect many important
features of a graph through a simple look at its spectral plot. We now
display some examples.\footnote{All networks are taken as undirected and unweighted. Thus, we suppress some potentially important aspects of the underlying data, but as our plots will show, we can still detect distinctive qualitative patterns. In fact, one can also compute the spectrum of directed and weighted networks, and doing that on our data will reveal further structures, but this is not explored in the present paper.}\\
First of all, the properties of the visualization will obviously depend on the display style, and this will be described first, see Figure \ref{MetabolicFigExmp}. That figure is based on the metabolic network of C.elegans. The first diagram displays the binned eigenvalues, that is, the range $[0,2]$ is divided here into 35 disjoint bins, and the number of eigenvalues that fall within each such bin is displayed, normalized by the total number of eigenvalues (relative frequency plot). The next figure smoothes this out by using overlapping bins, see figure legend for parameter values. The subsequent subfigures instead convolve the eigenvalues with a Gaussian kernel, that is, we plot the function
$$ f(x)=\sum_{\lambda_j}\frac{1}{\sqrt{2\pi\sigma^2}}\exp(-\frac{|x-\lambda_j|^2}{2\sigma^2}) $$
where the $\lambda_j$ are the eigenvalues. Smaller values of the variance $\sigma^2$ emphasize the finer details whereas larger values bring out the global pattern more conspicuously.\\
So, this is a network constructed from biological data. We next exhibit spectral plots, henceforth always with the value $\sigma=.03$, of networks constructed by formal schemes that have been suggested to capture important features of biological networks, namely an Erd\"os-R\'enyi random network, a Watts-Strogatz small-world network and a Barab\'asi-Albert scale-free network (see Figure \ref{GeneralModelsFigs}). It is directly obvious that these spectral plots are very different from the metabolic network. This suggests to us that any such generic network construction misses important features and properties of real biological networks. This will now be made more evident by displaying further examples of biological networks. We shall also see that biological networks from one given class typically have quite similarly looking spectral plots, which, however, are easily disitnguishable from those of networks from a different biological class. First, in Figure \ref{MetabolicFigs}, we show some further metabolic networks. Figures \ref{TranscripFigs} and \ref{PPINFigs} display transcription and protein-protein interaction networks. These still look somewhat similar to the metabolic networks, and this may reflect a common underlying principle. By way of contrast, the neurobiological networks in Figure \ref{NeuroFigs} and the food-webs in Figure \ref{FoodWebFigs} are entirely different -- which is not at all surprising as they come from different biological scales. \\\\
In conclusion, we have presented a simple technique for visualizing the important qualitative aspects of biological networks and for distinguishing networks from different origins. We expect that our method will further aid formal analysis of biological network data.

\end{document}